# Towards Expeditious and Unswerving Routing to Corroborate Nascent Internet

Dr Shishir Kumar, Mahesh kumar

**Abstract**—The internet is now-a-days experiencing a stress due to some inherent problems with the main interdomain routing protocol, boarder gateway protocol (BGP), the amount of time it takes to converge, number of update message exchanged followed by a failure to stabilize, the amount of time required to get a valid alternate path following the failure, the way size of routing table increasing, and security issues like integrity and privacy of routing tables and routing updates exchanged among the routers, are of our primary concern. In our proposed research work we plan to address aforementioned issues related to internet routing specially in boarder gateway protocol to enable BGP to offer expeditious unswerving routing to corroborate nascent internet. We plan to make some changes in the design of boarder gateway protocol and may introduce addition of extra features in BGP to help support above mentioned objective.

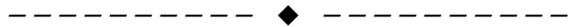

―――――――――  ◆  ―――――――――

## 1  INTRODUCTION

THE structure of the internet is a is a collection of networks, or Autonomous Systems (AS's), as shown in figure1 and 2, which are interconnected to form a connected domain [19]. Each AS uses an interior routing system to maintain a coherent view of the topology within the AS, and uses an exterior routing system to maintain adjacency information with neighboring AS's and thereby create a view of the connectivity of the entire system.

This network-wide connectivity is described in the routing table used by the BGP4 protocol. Each entry in the table refers to a distinct route. The attributes of the route are used to determine the best path from the local AS to the AS that is originating the route. Determining the 'best path' in this case is determining which routing advertisement and associated next hop address is the most preferred. The BGP routing system is not aware of finer level of topology within the local AS or within any remote AS. From this perspective BGP can be seen as a connectivity maintenance protocol, and the BGP routing table, a description of the current connectivity of the Internet, using an AS as the basic element of computation.

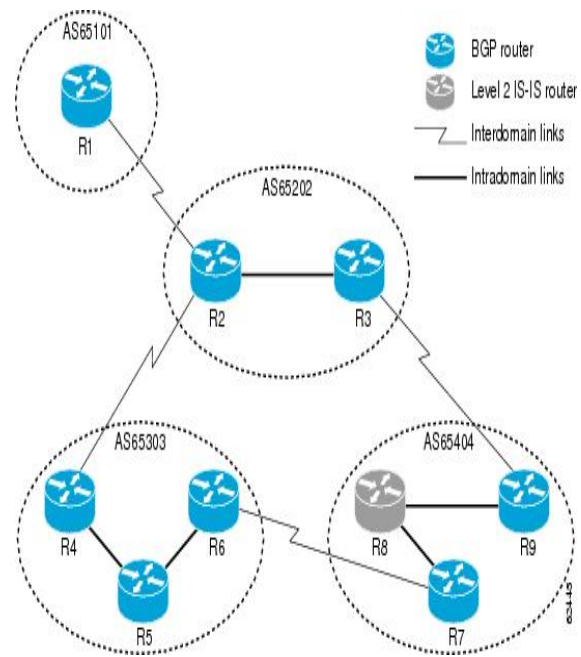

Figure 1: Autonomous Systems

――――――――――――――――――
• Dr Shishir Kumar is Head of the Department of Computer Science and Engineering, Jaypee Institute of Engineeringv and Technology, Guna, Madhya Pradesh (India).
• Mahesh Kumar is with the Department of Computer Science and Engineering, Jaypee Institute of Engineeringv and Technology, Guna, Madhya Pradesh (India).

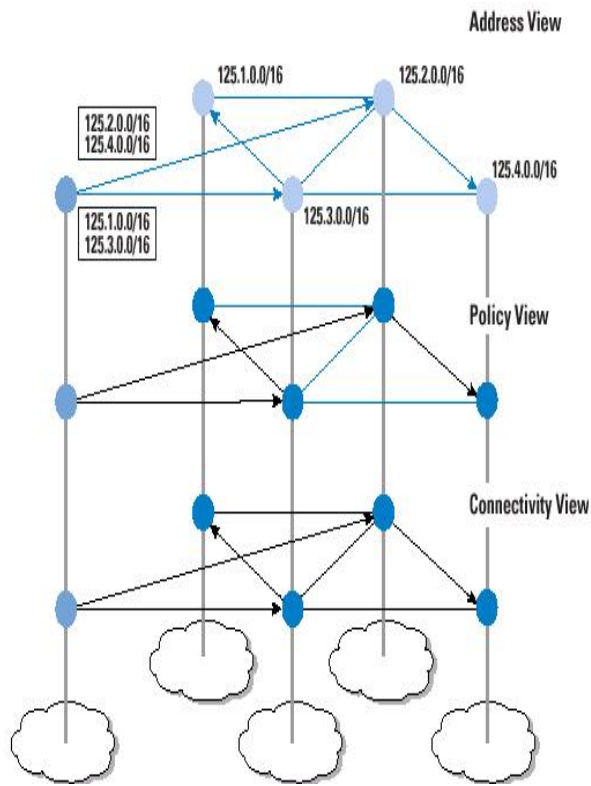

Figure 2: Multi-tier inter-domain routing

## 2 RELATED WORK:

Different authors have tried different approaches to overcome issues in internet routing problems, and each one has claimed to achieved better performance by either introducing new approaches or by making some different modifications to one factor or two of the existing protocol, but as we know that every factor is not completely unrelated to others, so by just ignoring the impact of change in one parameter of the protocol on others is not convincing.So It becomes necessary to compare all the results together and see that when one factor is reduced then its impact on all whole internet. Here its an attempt to study the wholesomeness internet performance and improve it.

## 3 CONVERGENCE TIME

A study of packet delivery performance during routing convergence have shown that network failures happen frequently, and that existing routing protocols converge slowly after a failure. During these routing convergence periods, some packets may already be enroute to their destinations and new packets may be sent. These on the way packets can encounter routing loops, **delays**, and losses [1]. Sometimes BGP takes a substantial amount of time and messages to converge and stabilize following the failure of some node in the Internet. A very common technique Route Flap Damping was introduced in BGP protocol to minimize the impact of relatively unstable routes, and almost all router manufacturers use this approach in their routers. Cisco and Juniper use in their routers to deliberately delay route calculations to increase stability. But flip side of RFD is the that sometime it can delay the network convergence, through simulation results it has been observed that route flap damping can significantly exacerbate the convergence times of relatively stable routes[3]. A new mechanism BGP-RCN, that provides an upper bound of O(d) on routing convergence delay for BGP, where d is the network diameter as measured by the number of AS hops. BGP-RCN lets each routing update message carry the information about the specific cause which triggered the update message. Once a node v receives the first update message triggered by a link failure, v can avoid using any paths that have been obsolete by the same failure [10].

## 4 SCALABILITY

Whenever we try to reduce the amount of time taken to converge, we try to keep most of the information about alternative paths to the destination in the router. But if we do not carefully look at the entries coming into the routing table after a failure or change in network, then routing table entries may increase to a point where it becomes difficult to manage it. Network operators and developers have shown their concern about the routing table growing at alarming rate which has potential to affect entire Internet. The Internet continues along a path of seemingly inexorable growth, at a rate that has, at a minimum, doubled in size each year. How big it needs to be to meet future demands remains an area of somewhat vague speculation. Of more direct interest is the question of whether the basic elements of the Internet can be extended to meet such levels of future demand, whatever they may be. To rephrase this question, are there inherent limitations in the technology of the Internet—or its architecture of deployment—that may impact the continued growth of the Internet to meet ever-expanding levels of demand [19]? The current Internet interdomain routing system will not be able to scale properly to meet growing needs, researchers have not yet produced any routing architecture with satisfactory approach to limit the growth of aforementioned entries. We believe that we will be facing in near future a challenge to scale the size of the Internet so as the size of routing table entries.

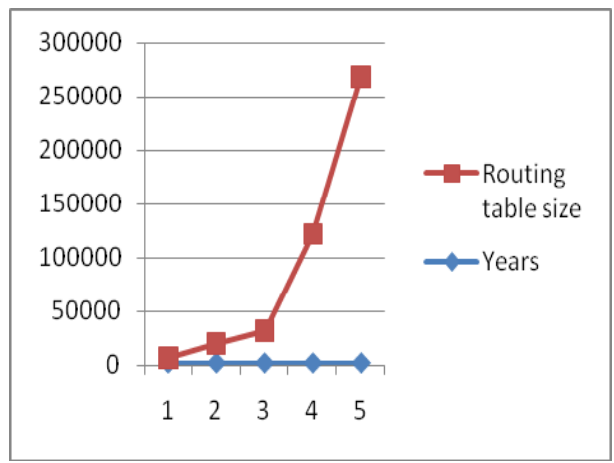

Figure 3: BGP Routing Table Growth Pattern

The last critical point was reached when the Internet's routing system adopted strong address aggregation using CIDR to handle address scaling in the mid 1990's. While CIDR was an extremely effective tactic, most experts agree that the growth behavior of the routing table in the last decade confirms that the type of address management required by CIDR will not suffice to meet future Internet routing needs [1]..

Today the growing Internet needs a reasonably good aggregation like CIDR has done last time which can reduce the size of routing table and change the way the routing table grow. But we its not so easy because aggregation beyond a limit can not help us as we know that it may raise issues load balancing and other traffic engineering. The problem with the Internet's routing architecture is how the interdomain routing protocol algorithm and its BGP implementation scale with the size of the network. Poor scaling of a routing algorithm expresses itself in terms of rapid rates of growth of the routing table size. Poor scaling of routing table sizes exacerbates convergence problem [5]. Not only is the communication overhead of BGP known to be exponential [1], but the BGP RT size also appears to grow exponentially [5].

## 5 FAULT MANAGEMENT

Even as recently as a decade ago, the failure of an Internet based system would have been a relatively minor annoyance. Today, however, such failures have an enormous cost, make news headlines, and, above all, have serious consequences on our society. The coming years will see the reach of the Internet extending wider than ever before, and together with this increasing reach will come a need for robust operation far more stringent than in the past. These dual trends, one a "technology push" toward pervasiveness driven by the integration of computing, wireless communication, and sensing technologies on small devices, and the other a demand pull driven by the use of networks as a critical component of the world's information systems, form the backdrop for my research[7].

Fault-tolerance is the ability to operate correctly under faulty conditions. These conditions in a network setting can include the failure of network components such as physical link failure, node failure, and switch failures. Components within nodes can become faulty. Fault instances can be permanent or transient, or a combination of these. Today we need fault tolerance in internet routing.

## 6 ROUTING POLICIES

The configurations of routing protocols determine how packets traverse each of these levels of Internet topology. A routing protocol is responsible for exchanging information about the state of the network and deciding which paths to use to reach every destination. The output of the routing protocol is a forwarding table. The primary role of a routing protocol is to detect and avoid failed links, but it also allows operators to express preferences for different paths to shape how traffic flows across a topology. Routing between domains is determined by policy. Each autonomous system can, based on configured policy, independently select routing information from its neighbouring autonomous systems, and selectively propagate this information. These policies are not expressed in terms of hop-distance to destinations [11]. Depending on how these policies are constructed, then, the resulting policy-based paths to destinations may incur more router level hops than shortest-router-hop path routing. Some paths may be preferred because they are lightly-used, cheaper, or more reliable. The preferences for different paths constitute routing policy, which ISP operators express in the configuration of routing protocols [8].

## 7 ROBUSTNESS AND SECURITY

Without introducing a foolproof security mechanism to protect routers and their routing tables it does not seem possible to maintain a correct flow of information among desired sources and destinations, extending the networks, and services without compromises. But we also understand that securing Internet routing is a challenging task. We need a flexible and scalable protocol and most importantly, a deployment strategy, since the Internet consists today of hundreds of thousands of routers and tens of thousands of independent networks [13]. We have seen from the previous experiences that a single malfunctioning router can poison the routing tables of many other routers on the network.

There various issues like prefix hijacking, unauthorized advertisement of IP prefixes, use of illegitimate paths, spamming, worms, trojan horses, password manipulating, and phishing which we wish to include in our research work to have reliable interdomain routing protocol.

## 8 PROPOSED METHODOLOGY

In our approach, we plan to achieve expeditious and unswerving routing R based of the following concept:

$$R=f(T,S,F,Se,Po)$$

Where: **T** is the convergence time of boarder gateway protocol,

**S** is the size of the routing table of boarder gateway protocol,

**F** is fault tolerance factor of boarder gateway protocol,

**Se** is the security factor of boarder gateway protocol, and

**Po** is the routing policy factor on boarder gateway protocol.

That is our proposed routing approach is to optimize each above mentioned factor of boarder gateway protocol to get expeditious and unswerving routing decisions.

To reduce the convergence time of boarder gateway protocol we plan to adopt a method to suggest some changes in design of boarder gateway protocol like treating the updates differently based on their nature and the very purpose. We plan to simulate the internet topology with many ISPs connected together and then by injecting faults with the help of a separate node into the simulated environment and will observe the its time of convergence, and then we will modify SSLD, classify the updates, MRAI timer mechanisms to get reduced convergence time. We will run simulations large number of time to get relatively more accuracy if we get expected result the changes will be accepted otherwise some other changes will be done in the design.

An approach to internet routing scalability and flexibility is to sepa rate the identification and locator of a node in the network. In an architecture where one label identifies a node and a different label indicates its location, topological changes will only change the locators which are assumed to follow topology and allow for aggregation, and then we plan to introduce an addition translation mechanism, after labelling a tag to each router belong to a particular area, which can translate it to normal identification which can be easily identified by router of different area.

Our proposed approach is to have a separate table of disjoint paths in a separate database in boarder gateway protocol for each possible source-destination pair if possible without strictly sticking to the condition of selecting shortest path.

Whenever a failure occurs in that situation the priority would be to get a new path from this database of disjoint paths table to forward traffic on alternate path.

To provide security and robustness to boarder gateway protocol we propose a strict method of content checking on each update received from neighbours, and if its content shows intuition of mal-intent then that update needs to be discarded and all the paths going through that router should be dropped from the database and a database of black listed router should be edited.

## 9 CONCLUSION

While studying the behaviour of boarder gateway protocol (BGP) we found that the convergence time, the number of update messages exchanged among routers, and the size of the routing tables increase at a rapid rate whenever network experience any change in the network connectivity/link failure. The entire network becomes over burdened and congested while processing these requests, we saw that sometime even a single route withdrawal message can trigger hundreds of route withdrawals and new route advertisements and the whole network of routers become overloaded with this work of stabilizing that user traffic has to suffer during this period.

In our proposed work we will limit the rate at which these withdrawal and advertisements emerge following a change in network connectivity so that the network can stabilize faster and will be available more to support user traffic. We also proposed methods to make BGP robust and secure by having a separate table of disjoint paths in a separate database for each possible source-destination pair if possible without strictly sticking to the condition of selecting shortest path. Whenever a failure occurs in that situation the priority would be to get a new path from this database of disjoint paths table to forward traffic on alternate path.

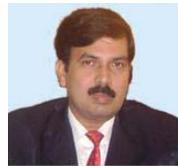

**Dr Shishir Kumar** is currently working as Associate Professor in the department of Computer Science and Engineering; He is also Head of the Computer Science and Engieering Department, in Jaypee Institute of Engineering and Technology, Guna, Madhya Pradesh, India. His academic qualification is Ph. D. (Computer Science) with more than 12 years of experience in rsearch and teaching.

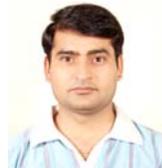

**Mahesh Kumar** is currently a lecturer in the department of Computer Science and Engineering in Jaypee Institute of Engineering and Technology, Guna, Madhya Pradesh, India, He has received is M. Tech. degree in Information Technology from Punjabi University, Patiala. He has more than 8 years of experience of managing computer networks (Cisco routers, switches, firewalls), and teaching.